\documentstyle[12pt]{article}

\def\dspace{\baselineskip = .30in}

\begin{document}

\title{The Stability of the Gauge Hierarchy
in $SU(5) \times SU(5)$ \footnote{Supported 
in part by Department of Energy grant
\#DE-FG02-91ER406267}}

\author{{\bf S.M. Barr}\\ 
Bartol Research Institute\\ University of Delaware\\
Newark, DE 19716}

\date{BA-96-31}
\maketitle

\begin{abstract}

It has been shown that the Dimopoulos-Wilczek (or missing-VEV)
mechanism for doublet-triplet splitting can be implemented
in $SU(5) \times SU(5)$ models, which requires no adjoint Higgs
fields. This is an advantage from the point of view of 
string theory construction. Here the stability of the gauge hierarchy is
examined in detail, and it is shown that it can be
guaranteed much more simply than in $SO(10)$. In fact a $Z_2$
symmetry ensures the stability of the DW form of the expectation
values to all orders in GUT-scale VEVs. It is also shown that models
based on $SO(10) \times SU(5)$ have the advantages of
$SU(5) \times SU(5)$ while permitting complete quark-lepton
unification as in $SO(10)$.

\dspace

\end{abstract}

\newpage

\section{Introduction}

The impressive unification of gauge couplings in the MSSM$^1$
is regarded by many as strong evidence for supersymmetric
grand unification. This has led to renewed interest in finding
satisfactory unification schemes and to attempts to derive such schemes
from string theory. Many gauge groups have been considered.
The simplest possibilities, at least at first sight, are
$SU(5)$ and $SO(10)$. An $SU(5)$-based theory is strongly suggested by the
unification of couplings and mass relations such as $m_b(M_{GUT})
= m_{\tau}(M_{GUT})$, while 
$SO(10)$ gives more complete quark-lepton unification.
However, both these groups require adjoint Higgs for symmetry
breaking, which is a drawback from the point of view of string theory.$^2$
Moreover, there is the problem of finding a satisfactory mechanism
for doublet-triplet splitting$^3$ in models based on these groups.

The only technically natural scheme for doublet-triplet splitting
in $SU(5)$ is the missing-partner mechanism$^4$, but this
requires the large Higgs representations ${\bf 75}$, and
${\bf 50} + \overline{{\bf 50}}$. (The sliding-singlet mechanism$^5$
is unstable to radiative corrections in $SU(5)$, though it is not
in certain larger groups.$^6$) 

In $SO(10)$ the only available scheme
for doublet-triplet splitting is the Dimopoulos-Wilczek (or
``missing VEV") mechanism.$^7$ It has been shown$^{8,9}$ that this can be
implemented in a technically natural way in $SO(10)$, but there are
difficulties which require relatively elaborate model-building
to solve. The trickiest of these is breaking the rank of $SO(10)$
down to 4 at a high scale. (This is necessary if righthanded 
neutrinos are to have a large mass.) The problem is that the sector which
breaks the rank (probably having to contain
at least a ${\bf 16} + \overline{{\bf 16}}$ of Higgs fields) will,
if it couples to the Dimopoulos-Wilczek (DW) sector, generally destabilize
the DW form of the VEVs needed to give the doublet-triplet splitting.
On the other hand, if these two sectors do not couple --- or couple
only weakly --- to each other, there arise goldstone or pseudo-goldstone
bosons that badly affect $\sin^2 \theta_W$. In Ref. 8 a way was
proposed to overcome this problem, involving a totally antisymmetric
interaction among three distinct adjoint Higgs fields, $\sum_{abc}
(\Omega_1)^a_b (\Omega_2)^b_c (\Omega_3)^c_a$. However, having three
distinct adjoint Higgs fields with different symmetry properties
may be difficult or impossible to achieve in string theory.$^2$

Flipped $SU(5) \times U(1)$ can be broken to the Standard
Model without adjoint Higgs, and also admits an extremely
elegant implementation of the missing-partner mechanism.
However, as this is not really a grand unified group it does
not explain the precise unification of gauge coupling that
has been seen. Moreover, it does not give such successful
relations as $m_b = m_{\tau}$ at the unification scale.

A very elegant possibility that preserves the good features of
$SU(5)$ but avoids the problems mentioned above is $SU(5)
\times SU(5)$, with the Standard Model group contained in the
``diagonal" $SU(5)$ subgroup.$^{10,11,12}$ This allows the symmetry
to be broken without adjoint Higgs. Instead, there are
Higgs in the $({\bf 5}, \overline{{\bf 5}})$ + h.c., which
under the diagonal subgroup decompose into ${\bf 24} + {\bf 1}$.
Barbieri, Dvali, and Strumia$^10$ have pointed out that
the Dimopoulos-Wilczek mechanism is simply implemented in
this group and in other groups of the form $G \times G$.
These have been studied in the context
of string theory by a number of groups$^{12}$ and seem very promising.$^2$

In this paper we examine in detail the question of
the stability of the gauge hierarchy in $SU(5) \times SU(5)$. We find that
the DW form of the vacuum expectation values 
can in a simple way be rendered
stable to all orders in GUT-scale VEVs 
by merely a $Z_2$ symmetry. Some of the difficulties
that exist in implementing the DW mechanism in $SO(10)$
are avoided. But $SU(5) \times SU(5)$ still does
not give the full quark-lepton unification that
is the most beautiful feature of $SO(10)$. We show
that $SO(10) \times SU(5)$ does this while still avoiding the problems
of $SO(10)$ itself.

\section{The Stability of the Hierarchy in $SU(5) \times SU(5)$}

We will discuss an $SU(5) \times SU(5)'$ model with Higgs fields
in the following representations: $H_I = ({\bf 5}, \overline{{\bf 5}})$,
$\overline{H}_I = (\overline{{\bf 5}}, {\bf 5})$, where $I = 1,2$, 
and $h = ({\bf 5}, {\bf 1})$, $\overline{h} = (\overline{{\bf 5}},
{\bf 1})$, $h' = ({\bf 1}, {\bf 5})$, $\overline{h}' = ({\bf 1},
\overline{{\bf 5}})$. Aside from some gauge singlets, these are the
only Higgs fields that will be needed to do all the symmetry breaking.

Consider, first, the following terms in the superpotential
that contain only $H_1$ and $\overline{H}_1$: 

\begin{equation}
W_1 = A_1 M {\rm Tr}(H_1 \overline{H}_1)
+ B_1 M^{-1} {\rm Tr}(H_1 \overline{H}_1 H_1 \overline{H}_1)
+ B'_1 M^{-1} [{\rm Tr}(H_1 \overline{H}_1)]^2.
\end{equation}

\noindent
Here ${\rm Tr}(H_1 \overline{H}_1 H_1 \overline{H}_1) \equiv
\sum (H_1)^{\alpha}_{\alpha^{'}} (\overline{H}_1)^{\alpha^{'}}_{\beta}
(H_1)^{\beta}_{\beta^{'}} (\overline{H}_1)^{\beta^{'}}_{\alpha}$.
(We use unprimed Greek indices for $SU(5)$ and primed for
$SU(5)'$.) The mass scale $M$ is assumed to be
of order $10^{16}$ GeV. It is trivial to see that one solution is

\begin{equation}
\langle (H_1)^{\alpha}_{\alpha^{'}} \rangle =
\langle (\overline{H}_1)^{\alpha^{'}}_{\alpha} \rangle
= \left( \begin{array}{ccccc}
a_1 & & & & \\ & a_1 & & & \\ & & a_1 & & \\ & & & 0 & \\
& & & & 0 \end{array} \right),
\end{equation}

\noindent
where $a_1 = \sqrt{\frac{-A_1}{6 B'_1 + 2 B_1}} M$. This is the
form required for the Dimopoulos-Wilczek mechanism for
doublet-triplet splitting. The two vanishing diagonal entries are
responsible (see below) for the lightness of the two doublet
Higgs fields of the Standard Model. It should be noted that
the diagonal forms with $n$ vanishing
diagonal entries and $5-n$ equal to $a_1$ are also solutions.

This so far only breaks $SU(5) \times SU(5)'$ 
to $SU(3) \times SU(2) \times SU(2) \times U(1)$. The rest
of the breaking can be done by the fields $H_2$ and
$\overline{H}_2$. Assume that these have a superpotential of the same form as 
Eq. (1) (just replacing the index `1' everywhere in Eq. (1) with 
a `2') and from it acquire the VEVs

\begin{equation}
\langle (H_2)^{\alpha}_{\alpha^{'}} \rangle =
\langle (\overline{H}_2)^{\alpha^{'}}_{\alpha} \rangle
= \left( \begin{array}{ccccc}
0 & & & & \\ & 0 & & & \\ & & 0 & & \\ & & & a_2 & \\
& & & & a_2 \end{array} \right),
\end{equation}

\noindent
where $a_2 = \sqrt{\frac{-A_2}{6B_2 + 2B'_2}} M$. Taken together
the VEVs in Eqs. (2) and (3) break $SU(5) \times SU(5)'$
all the way down to the Standard Model. Henceforth, we will
call the form of the VEVs given in Eqs. (2) and (3) the ``DW form".

The two sectors, $(H_1, \overline{H}_1)$ and $(H_2, \overline{H}_2)$,
must couple together if goldstone modes are to be avoided. In 
particular, the generators of $\frac{SU(5)}{SU(3)\times SU(2)} 
\times \frac{SU(5)'}{SU(3)' \times SU(2)'}$ are broken in {\it both} sectors,
so that there are two sets of uneaten goldstones in the representations
$[(3,2)^{-\frac{5}{6}} + (\overline{3}, 2)^{\frac{5}{6}}]$ as well as
the two sets in $[(3,2)^{-\frac{5}{6}} + (\overline{3}, 2)^{\frac{5}{6}}]$
that do get eaten (one set for each $SU(5)$). 

The two sectors could be coupled together by a term like
${\rm Tr}(H_2 \overline{H}_1)$, but that would clearly
destabilize the DW form of the VEVs, as then, for instance, the
$F_{\overline{H}_1} = 0$ equation would have a term proportional
to $H_2$. Such a term must be ruled out, and this can be done
by a $Z_2$ symmetry, $K$, under which $H_2 \rightarrow
- H_2$ and $\overline{H}_2 \rightarrow - \overline{H}_2$. This
still allows the following mixed terms at the quartic level:

\begin{equation}
\begin{array}{ll}
{\rm Tr}(H_1 \overline{H}_1){\rm Tr}(H_2 \overline{H}_2), &
{\rm Tr}(H_1 \overline{H}_2){\rm Tr}(H_2 \overline{H}_1) \\
\\
({\rm Tr}(H_1 \overline{H}_2))^2, &
({\rm Tr}(H_2 \overline{H}_1))^2 \\
\\
{\rm Tr}(H_1 \overline{H}_1 H_2 \overline{H}_2), &
{\rm Tr}(H_1 \overline{H}_2 H_2 \overline{H}_1) \\
\\
{\rm Tr}(H_1 \overline{H}_2 H_1 \overline{H}_2), &
{\rm Tr}(H_2 \overline{H}_1 H_2 \overline{H}_1). \\
\end{array}
\end{equation}

\noindent
It is easy to check that the last four terms in this list
give mass to all of the would-be goldstone bosons that
are not eaten by the gauge bosons. For example, the term

\begin{equation}
(H_1)^i_{a'} \langle (\overline{H}_1)^{a'}_a \rangle
(H_2)^a_{i'} \langle (\overline{H}_2)^{i'}_i \rangle,
\end{equation}

\noindent
where we use $a$ for $SU(3)$ indices and $i$ for $SU(2)$ indices,
couples a $(\overline{3}, 2)^{\frac{5}{6}}$ in $H_1$
to a $(3,2)^{-\frac{5}{6}}$ in $H_2$. 

A crucial issue is whether the DW form is stable under the influence
of these operators. It is easy to see that it is. 
For example, it is necessary for the stability
of the DW form of the VEV of $H_1$ that there be no
contributions to the $F_{\overline{H}_1}$
that are proportional to powers of $H_2$ or $\overline{H}_2$.
The first term in the list gives a contribution to $F_{\overline{H}_1}$
of $H_1 {\rm Tr}(H_2 \overline{H}_2)$
which is proportional to $H_1$ as required. The second term in the list
contributes $H_2 {\rm Tr}(H_1 \overline{H}_2)$ to $F_{\overline{H}_1}$,
which might seem to be a problem, except that ${\rm Tr}(H_1 \overline{H}_2)$
vanishes when the fields take the forms given in Eqs. (2) and (3).
Similarly, the other mixed quartic terms listed above do not destabilize
the DW form.

Since essential use is being made of non-renormalizable
operators here (see the discussion of this below) it is important
that one show that the gauge hierarchy is stable even when
higher-than-quartic operators are taken into account. The zeros
on the diagonal of the DW form must vanish at least to
order $M_{GUT}^5/M_{Pl}^4$ and probably higher.

In fact, it is not difficult to show that the DW form given in
Eqs. (2) and (3) is stable {\it to all orders} in GUT-scale VEVs
because of the
simple $Z_2$ symmetry that we called $K$. Let us call the value
of some product of fields, $\Pi$, when these fields take
the DW form $\langle \Pi \rangle_{DW}$. Suppose there is a term,
$T$, in the superpotential $W$ that destabilizes the DW form
of the VEV of $H_1$. Then this term forces $\langle (H_1)^4_4 \rangle
= \langle (H_1) \rangle^5_5 \neq 0$. (Given the forms of $\langle
H_2 \rangle$ and $\langle \overline{H}_2 \rangle$, this is the only
potential instability of $H_1$ which we need consider.)
This could happen only if $\langle ( \partial T/ \partial 
\overline{H}_1)^{\alpha}_{\alpha'} \rangle_{DW} = t (\delta^{\alpha}_4
\delta^4_{\alpha'} + \delta^{\alpha}_5
\delta^5_{\alpha'}) \neq 0$. But that would imply that 
$\Sigma_{\alpha, \alpha'} \langle (\overline{H}_2)^{\alpha'}_{\alpha}
(\partial T/ \partial \overline{H}_1)^{\alpha}_{\alpha'} \rangle 
\neq 0$. In other words, replacing a factor of $\overline{H}_1$ by
$\overline{H}_2$ in $T$ leads to a term $T'$ which has the
property that $\langle T' \rangle_{DW} \neq 0$. In the same way
if the DW form of $H_2$ is unstable, there must be a term which
after a factor of $\overline{H}_2$ is replaced by $\overline{H}_1$
does not vanish when the fields take the DW form. 

If we can prove, then, that for {\it every} possible term in $W$, changing
one factor of $H_1$ (or $\overline{H}_1$) to $H_2$ (or $\overline{H}_2$) or
{\it vice versa} leads to a term $T'$ such that $\langle T' 
\rangle_{DW} = 0$, we shall have shown that the DW form is stable.
But the terms in $W$ all have, by the symmetry $K$, an even
number of factors of $H_2$ or $\overline{H}_2$. Thus replacing
a field with label 1 by one with label 2, or {\it vice versa},
always produces a terms with an odd number of factors
of $H_2$ or $\overline{H}_2$. All that needs to be shown, then,
is that for any expression, $T'$, with an odd number of factors of $H_2$
or $\overline{H}_2$, $\langle T' \rangle_{DW} = 0$. This is easy
to do.

Consider, first, a term, $T'$, with no $SU(5)$ or $SU(5)'$
$\epsilon$-symbols. Such a term must be a product of
factors of the form ${\rm Tr}(H_I \overline{H}_J H_K \cdot
\cdot \cdot \overline{H}_L)$. But if $\langle 
{\rm Tr}(H_I \overline{H}_J H_K \cdot
\cdot \cdot \overline{H}_L) \rangle_{DW} \neq 0$, then all
of the labels $I$, $J$, $K$, ..., $L$ must be the same, all
1's or all 2's --- and there are an even number of such labels.
Thus, if $\langle T' \rangle_{DW} \neq 0$ it must have
an even number of factors of $H_2$ or $\overline{H}_2$.

The argument is almost as simple for terms which contain
$SU(5)$ and $SU(5)'$ $\epsilon$-symbols. Consider a
term, $T'$ for which $\langle T' \rangle_{DW} \neq 0$.
The simplest thing is
to keep track of the $SU(5)$ ($SU(5)'$) indices that
take values in the $SU(2)$ ($SU(2)'$) subgroup. Each
$\epsilon$-symbol has two such indices, and three that
take values in the $SU(3)$ ($SU(3)'$) group. Because
$\langle T' \rangle_{DW} \neq 0$ it must be that
each factor of $H_2$ or $\overline{H}_2$ has one $SU(2)$
and one $SU(2)'$ index. And each factor of $H_1$ or
$\overline{H}_1$ has no $SU(2)$ or $SU(2)'$ indices.
Thus if we think of an $SU(2)$ or $SU(2)'$ index as
being a line, we can think of $\epsilon$-symbols and
factors of $H_2$ and $\overline{H}_2$ as being
vertices into which precisely two lines enter, and there
are no n-vertices with $n \neq 2$. Thus the lines representing
the $SU(2)^{(')}$ indices go around in a loop, and the
term $T'$ must in such a diagram be represented by a
set of disconnected single loops. Now in each such
loop $SU(2)$ indices are converted into $SU(2)'$
indices, or {\it vice versa}, by the factors of
$H_2$ and $\overline{H}_2$, but {\it not} by the
$\epsilon$-symbols. Thus each loop must contain an {\it even}
number of factors of $H_2$ or $\overline{H}_2$ and so also,
therefore, must the term $T'$. But this is what we had to prove.
Thus the VEVs given in Eqs. (2) and (3) are stable
solutions to all orders because of the simple $Z_2$
symmetry, $K$. 

The only other Higgs fields in the model, $h$, $\overline{h}$,
$h'$, and $\overline{h}'$, have either vanishing or Weak-scale
VEVs, so that the stability of the gauge hierarchy is guaranteed by what has
been said already.

In the foregoing, we have made essential use of non-renormalizable
operators, and the question arises whether these could have been
the result of integrating out fields of mass $M \sim M_{GUT}$.
The answer is yes if the fields integrated out include adjoints
of $SU(5)$ and $SU(5)'$. For example, if one had $H_1 \overline{H}_2 X
+ MX^2$, integrating out $X$ would give $(H_1 \overline{H}_2)
(H_1 \overline{H}_2)$. But if $X$ is a singlet, then only the
contraction $[{\rm Tr}(H_1 \overline{H}_2)]^2$ results. This does
not give mass to the uneaten would-be goldstone bosons. If $X$
is a $({\bf 24}, {\bf 24})$ then one obtains Tr$(H_1 \overline{H}_2
H_1 \overline{H}_2)$ as required to give mass to the goldstones.
Note as well that $\langle X \rangle$ is determined by the
equation $F_X = 0$ to be $\langle X \rangle = \frac{1}{2} \langle H_1
\rangle \; \langle \overline{H}_2 \rangle = 0$. Thus the desired form
of Eqs. (2) and (3) is not destabilized by $\langle X \rangle$.
The choice would therefore seem to be between having adjoint
representations of $SU(5)$ or having higher-dimensional operators
as in Eq. (1).

The $K$-invariant operators involving $H_1$, $\overline{H}_1$,
$H_2$, and $\overline{H}_2$ up to fourth order have an accidental
$U(1)^2$ symmetry. The charges of $(H_1, \overline{H}_1,
H_2, \overline{H}_2)$ under these $U(1)$'s are $(1,1,-1,-1)$ and
$(1,-1, 1,-1)$. One might worry, therefore,
about axions or goldstone bosons. But at quintic and higher order
there are $K$-invariant operators that explicitly break these
accidental $U(1)$'s. For example, there is 
$(H_1)^{\alpha}_{\alpha'}
(H_1)^{\beta}_{\beta'} (H_1)^{\gamma}_{\gamma'} (H_1)^{\delta}_{\delta'}
(H_1)^{\epsilon}_{\epsilon'} \epsilon_{\alpha \beta \gamma \delta
\epsilon} \epsilon^{\alpha' \beta' \gamma' \delta' \epsilon'}$ and the
same structure with two or four factors of $H_1$ replaced by
$H_2$. 

It is an interesting question 
whether other possibilities exist for breaking down
to the Standard Model gauge group besides the pattern of VEVs given
in Eqs. (2) and (3). It was noted in Ref. 9 that any two of the
three forms, diag$(a,a,a,0,0)$, diag$(0,0,a,a,a)$, and diag$(a,a,a,a,a)$,
will achieve this breaking. However, the inclusion of VEVs
in the $SU(5)_V$-singlet direction diag$(a,a,a,a,a)$
seems to greatly complicate the problem of achieving a stable
DW mechanism. For example let there be a set of Higgs,
$H_3$ and $\overline{H}_3$, with VEVs in this direction as
well as the set $H_1$ and $\overline{H}_1$. In this case
the generators in $\frac{SU(3) \times SU(3)'}{SU(3)_V}$ are broken
in both the $(H_1, \overline{H}_1)$ and $(H_3, \overline{H}_3)$ 
sectors. To avoid these (and other) consequent goldstone modes these
two sectors must be coupled. However, the method that worked
above does not work here. A $Z_2$ symmetry under which $H_3$
and $\overline{H}_3$ are odd will indeed forbid the dangerous
term Tr$(H_3 \overline{H}_1)$, but terms like $[{\rm Tr}
(H_3 \overline{H}_1)]^2$ are no less dangerous, since 
Tr$(H_3 \overline{H}_1)$ does not vanish, and therefore all such
mixed terms destabilize the hierarchy. It may be possible to
stabilize the hierarchy in the presence of such $SU(5)_V$-singlet
VEVs, but it would doubtless be complicated. This problem
is reminiscent of the difficulty of stabilizing the DW form in
$SO(10)$ in the presence of $SU(5)$-singlet VEVs which are
needed there to break the rank of the group and make
righthanded neutrinos superheavy.$^{8,9}$ 

So far it has been shown that one can achieve the DW form (Eq. (2)),
have it stable to all orders in the GUT-scale VEVs, completely
break to the Standard Model gauge group, and avoid pernicious
goldstone modes. All of this we have achieved with a simple $Z_2$
discrete symmetry and a small set of Higgs representations.
This is quite simple compared to what was shown 
to be necessary in $SO(10)$ in Refs. 8 and 9.

The doublet-triplet splitting itself can be achieved in a way closely
analogous to what was suggested in Refs. 8 and 9 for $SO(10)$.
Gauge symmetry allows the terms $M \overline{h} h$
and $M' \overline{h}' h'$. If both of these are present
with $M \sim M' \sim M_{GUT}$, then all of the Weak-doublet
Higgs become superheavy and the gauge hierarchy is destroyed.
Suppose, then, that a discrete symmetry, $K_M$, forbids
the $M \overline{h} h$ term, but allows $M' \overline{h}' h'$.
Then the pair of doublets in $\overline{h} + h$ would remain light,
to play the role of the Higgs of the MSSM. The color-triplets
in $\overline{h} + h$ become heavy from the terms $\overline{h}
\langle H_1 \rangle h'$ and $\overline{h}' \langle \overline{H}_1 
\rangle h$. The triplets have a $2 \times 2$ mass matrix of the form

\begin{equation}
(\overline{h}_a, \overline{h}'_{a'})
\left( \begin{array}{cc} 0 & \langle (H_1)^a_{a'} \rangle \\
\langle (\overline{H}_1)^{a'}_a \rangle & M'
\end{array} \right) \left( \begin{array}{c}
h^a \\ h'^{a'} \end{array} \right),
\end{equation}

\noindent
while the doublets have the mass matrix

\begin{equation}
(\overline{h}_i, \overline{h}'_{i'})
\left( \begin{array}{cc} 0 & \langle (H_1)^i_{i'} \rangle =0 \\
\langle (\overline{H}_1)^{i'}_i \rangle = 0 & M'
\end{array} \right) \left( \begin{array}{c}
h^i \\ h'^{i'} \end{array} \right).
\end{equation}

\noindent
In order for the doublets in $\overline{h} + h$ to be light,
there must be no couplings $\overline{h} H_2 h'$ and 
$\overline{h}' \overline{H}_2 h$. But this is insured by
just the $Z_2$ symmetry $K$ which reflects $H_2$ and $\overline{H}_2$,
as long as $h$, $\overline{h}$, $h'$, and $\overline{h}'$
transform trivially under it. Moreover, such allowed 
higher-dimension operators as $(\overline{h} H_2 h')
{\rm Tr}(H_1 \overline{H}_2)$ are not dangerous because
of the vanishing of ${\rm Tr}(H_1 \overline{H}_2)$ at the
DW minimum. It is clear that the symmetry $K$ together with
the DW forms of the VEVs protect the hierarchy from such
operators to all orders.

The only non-trivial problem is to ensure to sufficiently
high order to preserve the gauge hierarchy
that there is no term
that effectively gives $M \overline{h} h$. However,
it was shown how to solve this problem for $SO(10)$ in Ref. 9,
and the same kinds of solutions work here as well. A simple
possibility is the following. Let $K_M = Z_n$. Suppose
there is a singlet Higgs field, S, which under $K_M$ 
transforms as $S \rightarrow z \; S$ (but that
there is {\it no} singlet filed $\overline{S}$ that transforms as
$\overline{S} \rightarrow z^* \overline{S}$). And suppose that
under $K_M$ one has $h \rightarrow z \; h$, $\overline{h}' \rightarrow
z^* \overline{h}'$, with all other fields transforming
trivially. Then $K_M$ allows $S \overline{h}' h'$,
$\overline{h} H_1 h'$, and $\overline{h}' \overline{H}_1 h$,
but forbids $\overline{h} h$ and $S \overline{h} h$. The 
lowest dimension operator that gives mass to the light doublets is
then $S^{n-1} \overline{h} h/M_{Pl}^{n-2}$. Thus the hierarchy can be made
stable enough by making $n$ large. Of course, there are other,
and perhaps more elegant, possibilities for $K_M$, but this example 
is enough to show that the dangerous term $\overline{h} h$
can be sufficiently suppressed. (For a more complete discussion
of the problem see Ref. 9.)

In the scheme for doublet-triplet splitting just described,
the amplitude for Higgsino-mediated proton decay is suppressed by 
a factor of $M'/M_{GUT}$. As noted in Ref. 8 it would be
possible with a VEV of the form given in Eq. (3) replacing
the mass $M'$ to suppress proton decay completely. 
However, it does not seem possible in the present
context to have a term like $\overline{h}' \overline{H}_2 H_2 h'$
(which respects $K$) without also having $M' \overline{h}' h'$.

The light quarks and leptons
most simply come from $\psi^A_{10} = ({\bf 10}, {\bf 1})$
and $\psi^A_{\overline{5}} = ( \overline{{\bf 5}}, {\bf 1})$
of $SU(5) \times SU(5)'$, with $A = 1,2,3$ being the family
index. Then $\psi^A_{10} \psi^B_{10} h$ and $\psi^A_{10}
\psi^B_{\overline{5}} \overline{h}$ give quark and lepton
masses just as in minimal $SU(5)$. 

If $a_1$ and $a_2$ (in the VEVs of $H_1$ and $H_2$) are very
different, then the value of $\sin^2 \theta_W$ would differ
substantially from the minimal SUSY $SU(5)$ value. For
example, if $a_1 \gg a_2$, one would have $SU(5) \times
SU(5)' \rightarrow SU(3)_V \times SU(2) \times SU(2)' 
\times U(1)_V \rightarrow SU(3)_V \times SU(2)_V \times
U(1)_V$, and corrections to $\sin^2 \theta_W$ would
be of order $\frac{\alpha}{30 \pi} \ln(a_1/a_2)$. We can crudely
estimate the effect on $\sin^2 \theta_W$ as follows. The
largest threshold corrections come from multiplets in
$({\bf 8}, {\bf 1})^0$ and $({\bf 1}, {\bf 3})^0$ of
the Standard Model group, which contribute at one-loop
$\Delta \sin^2 \theta_W (M_Z)  = \frac{\alpha}{30 \pi}
[21 g_8 \ln M_8 - 24 g_3 \ln M_3]$. ($g_8$ and $g_3$ represent the
effective number of Higgs multiplets in the representations
$({\bf 8}, {\bf 1})^0$ and $({\bf 1}, {\bf 3})^0$, and
$M_8$ and $M_3$ represent their masses.) It is
easily seen that $M_8 \sim a_1$ and $M_3 \sim a_2$. Moreover,
$g_8 = g_3$. Thus one expects an effect $\Delta \sin^2
\theta_W \sim g \frac{\alpha}{30 \pi} 21 \ln (a_1/a_2)
\sim 10^{-2} \ln (a_1/a_2)$, if $a_1/a_2$ is very different
from unity. However, one expects $a_1/a_2$ to be of order
unity, and for that case the different multiplets will
contribute with varying signs to the threshold corrections,
which one would therefore expect to be somewhat less than $10^{-2}$.

\section{$SO(10) \times SU(5)$}

The group $SU(5) \times SU(5)$ has been shown to have
two main advantages over $SO(10)$, namely the possibility
of breaking symmetry all the way to $SU(3) \times SU(2)
\times U(1)$ without adjoint Higgs fields, and
the simplicity of preserving the hierarchy while doing so.
However, $SO(10)$ has at least one greatly attractive
feature, which is that it achieves complete quark-lepton
unification. All the fermions of a generation are
unified into one irreducible representation, and therefore
the up-quark masses are related to the masses of the
down-quarks and charged leptons. Moreover, righthanded
neutrinos are predicted to exist.

If one generalizes the discussion given in the previous
section to the group $SO(10) \times SU(5)$ one easily
sees that one can have all the advantages of $SU(5) \times
SU(5)$ combined with those of $SO(10)$.

In an $SO(10) \times SU(5)'$ model let there be the following
Higgs fields: $H_I = ({\bf 10}, \overline{{\bf 5}})$, $\overline{H}_I
= ({\bf 10}, {\bf 5})$, $I = 1,2$, and $h = ({\bf 10},
{\bf 1})$, $h' = ({\bf 1}, {\bf 5})$, and $\overline{h}' =
({\bf 1}, \overline{{\bf 5}})$. The discussion closely
parallels that in the last section. $SO(10) \times SU(5)'$
contains $SU(5) \times SU(5)'$ as a subgroup, under which
$H_1$ decomposes to a $({\bf 5}, \overline{{\bf 5}})$ (which
just corresponds to the $H_1$ of the last section) and a
$(\overline{{\bf 5}}, \overline{{\bf 5}})$. With a superpotential
that contains at least up to quartic terms, one can ensure that
the $({\bf 5}, \overline{{\bf 5}})$ parts of $H_I$ have VEVs of the
same forms shown in Eqs. (2) and (3), and similarly for the 
$\overline{H}_I$. These VEVs are enough to break the group down to
that of the Standard Model. It can be shown that quartic terms also
give mass to
all would-be goldstones as before. (There are more quartic terms here,
as for $SO(10) \times SU(5)$ the indices can be contracted
in more ways.) For example, in $SU(5) \times SU(5)'$
notation, terms like $H_{\alpha \alpha'} \langle 
\overline{H}^{\alpha'}_{\beta} \rangle H^{\beta \beta'}
\langle \overline{H}^{\alpha}_{\beta'} \rangle$ give
mass to goldstones which are in ${\bf 10} + \overline{{\bf 10}}$
representation of $SU(5)$, that is, the ones from the
coset $\frac{SO(10)}{SU(5)}$.

The DW forms in Eqs. (2) and (3) can be guaranteed to all
orders in GUT-scale VEVs by the same $Z_2$ symmetry
as in the last section. Since the $SU(5) \times SU(5)'$ $({\bf 5},
{\bf 5})$ and $(\overline{{\bf 5}}, \overline{{\bf 5}})$
parts of $H_I$ and $\overline{H}_I$ get no VEVs they can
be ignored, and the proof of stability reduces exactly to that
in the last section.

The doublet-triplet splitting is achieved just as
in the last section, the only difference being that
what we called $h$ and $\overline{h}$ there are both
contained in the ten-dimensional $h$ in the present case.
Thus the term that must be forbidden by ``$K_M$" is $h^2$.

The quark-lepton unification is achieved simply by
putting the known fermions in $({\bf 16}, {\bf 1})$
representations, which we shall call $\psi^A_{16}$,
where $A$ is the family index. Then all the Dirac
mass matrices come from the coupling $\psi^A_{16} \psi^B_{16} h$,
and one has the possibility of relating the masses of
the up quarks to those of the down quarks and leptons
that is an important feature of $SO(10)$.

A possible difficulty is the generation of the Majorana
masses of the righthanded neutrinos. These require a product of
VEVs that is in a $({\bf 126}, {\bf 1})$.
A ${\bf 126}$ field cannot be constructed in $SO(10)$ string
models$^2$, but one can get this effectively from
a product of two spinor Higgs fields: $(\overline{{\bf 16}}, {\bf 1})^2$.
The problem, however, is that that raises all the difficulties
in preserving the stability of the gauge hierarchy that one
encounters in $SO(10)$ models.$^{8,9}$
The $SU(5)$-singlet VEV will either lead to unwanted
goldstones (if the spinor-Higgs decouples from the
DW sector) or will destabilize the DW form (if the sectors are
coupled). The trick that allowed a resolution of this
difficulty in Ref. 8 involved the existence of three
adjoint Higgs fields, which we are eschewing in the present
approach. 

Fortunately, however, one need introduce no additional
fields in this $SO(10) \times SU(5)$ model
to generate large righthanded neutrino masses.
The following term suffices:

\begin{equation}
{\cal O} = \psi^A \Gamma^{abcde} \psi^B (H_1)^a_{\alpha}
(H_1)^b_{\beta} (H_1)^c_{\gamma} (H_2)^d_{\delta} (H_2)^e_{\epsilon}
\epsilon^{\alpha \beta \gamma \delta \epsilon}/{\cal M}^4.
\end{equation}

\noindent
Note that the product of $H_I$ fields is totally
antisymmetric under both $SO(10)$ and $SU(5)'$, and
so symmetric in Higgs ``flavor". Note also that
this term is even under the $Z_2$ symmetry. The five Higgs
are contracted precisely into a $({\bf 126}, {\bf 1})$
and, with the VEVs in Eqs. (2) and (3), give a mass
of order $M_{GUT}^5/{\cal M}^4$ to the righthanded 
neutrinos. If we assume that ${\cal M}$ is about $10$ to
$20$ times $M_{GUT}$ this gives $M_R$ at an intermediate
scale of $10^{11}$ to $10^{12}$ GeV.

\section*{Acknowledgments}

The author thanks K.S. Babu for very helpful comments.

\section*{References}

\begin{enumerate}

\item U. Amaldi, W. deBoer, and H. Furstenau, Phys. Lett. {\bf 200B},
447 (1991); P. Langacker and M.X. Luo, Phys. Rev. {\bf D44}, 817 (1991).
\item For a review of string theory constructions of unified theories
see K.R. Dienes, IASSNS-HEP-95/97, hep-th/9602045, to appear in
Physics Reports. For recent results on allowed representations
in string-derived $SO(10)$ models see K.R. Dienes, IASSNS-HEP-96/64,
hep-ph/9606467.
\item For a recent review of the doublet-triplet splitting
problem, see L. Randall, Talk presented at
SUSY '95 Conference, hep-ph/9508208.
\item A. Buras, J. Ellis, M. Gaillard, and D.V. Nanopoulos, Nucl. Phys.
{\bf B135}, 66 (1985); H. Georgi, Phys. Lett. {\bf 108B}, 283 (1981);
A. Masiero, D.V. Nanopoulos, K. Tamvakis, and T. Yanagida, Phys. Lett.
{\bf 115B}, 380 (1982); B. Grinstein, Nucl. Phys. {\bf B206}, 387 (1982).
\item E. Witten, Phys. Lett. {\bf 105B}, 267 (1981); D.V. Nanopoulos,
and K. Tamvakis, {\it ibid.} {\bf 113B}, 151 (1982); S. Dimopoulos
and H. Georgi, {\it ibid.} {\bf 117B}, 287 (1982); L. Ibanez and G. Ross,
{\it ibid.} {\bf 110B}, 215 (1982).
\item A. Sen, Phys. Lett. {\bf 148B}, 65 (1984).
\item S. Dimopoulos and F. Wilczek, NSF-ITP-82-07 (1981).
R.N. Cahn, I. Hinchliffe, and L. Hall, Phys. Lett. {\bf 109B}, 426 (1982).
\item K.S. Babu and S.M. Barr, Phys. Rev. {D48}, 5354 (1993).
\item K.S. Babu and S.M. Barr, Phys. Rev. 
{\bf D50}, 3529 (1994); {\it ibid.} {\bf D51}, 2463 (1995).
\item R. Barbieri, G. Dvali, and A. Strumia, Phys. Lett. {\bf 333B},
79 (1994).
\item R.N. Mohapatra, hep-th/9601203.
\item G. Aldazabal, A. Font, L.E. Ibanez, and A.M. Uranga, Nucl. Phys.
{\bf B452}, 3 (1995); hep-ph/9410206; hep-ph/9508033;
S. Chaudhuri, G. Hockney, and J. Lykken, hep-ph/9510241.

\end{enumerate}

\end{document}